\documentclass{raa} 
\usepackage{graphicx,url,longtable,natbib} 

\newcommand{\kms}{\,\rm{km\,s^{-1}}}
\newcommand{\dv}{|\Delta V|}
\newcommand{\rp}{r_{\rm p}}
\newcommand{\kpc}{\,h_{70}^{-1} {\rm kpc}}
\newcommand{\Mflag}{\rm{M\_Flag}}
\newcommand{\Oflag}{\rm{O\_Flag}}

\begin{document}

\title{A sample of galaxy pairs identified from the LAMOST spectral survey and
the Sloan Digital Sky Survey}
\volnopage{Vol.0 (200x) No.0, 000--000}
\author{Shiyin Shen \inst{1,2}
\and M. Argudo-Fern\'andez \inst{1}
\and Li Chen \inst{1}
\and Xiaoyan Chen \inst{3}
\and Shuai Feng \inst{1,4}
\and Jinliang Hou \inst{1}
\and Yonghui Hou \inst{5}
\and Peng Jiang \inst{6}
\and Yipeng Jing \inst{7}
\and Xu Kong \inst{8}
\and Ali Luo  \inst{3}
\and Zhijian Luo\inst{2}
\and Zhengyi Shao \inst{1,3}
\and  Tinggui Wang  \inst{6}
\and Wenting Wang \inst{8}
\and Yuefei Wang \inst{5}
\and Hong Wu \inst{3}
\and Xue-Bing Wu \inst{9,10}
\and  Haifeng Yang \inst{3}
\and  Ming Yang \inst{3}
\and Fangting Yuan\inst{1,2}
\and Hailong Yuan\inst{3}
\and Haotong Zhang\inst{3}
\and Jiannan Zhang  \inst{3}
\and Yong Zhang \inst{5}
\and Jing Zhong \inst{1}
 }

\institute{Key Laboratory for Research in Galaxies and Cosmology, Shanghai
Astronomical
Observatory, Chinese Academy of Sciences, 80 Nandan Road, Shanghai,
200030, China, {\it ssy@shao.ac.cn};\\
\and Key Lab for Astrophysics, Shanghai, 200234, China;\\
\and National Astronomical Observatories, Chinese Academy of Sciences, Beijing,
100012, China;\\
\and University of Chinese Academy of Sciences, Beijing, 100049, China;\\
\and Nanjing Institute of Astronomical Optics \& Technology, National
Astronomical Observatories, Chinese Academy of Sciences, Nanjing 210042, China
\\
\and University of Science and Technology of China, Hefei, 230026, China;\\
\and Department of Physics and Astronomy, Shanghai Jiao Tong University,
Shanghai, 200240, China; \\
\and Institute for Computational Cosmology, University of Durham, South Road,
Durham, DH1 3LE, UK \\
\and Department of Astronomy, Peking University, Beijing, 100871, China;\\
\and Kavli Institute for Astronomy and Astrophysics, Peking University, Beijing
100871, China\\
}

\abstract{A small fraction($<10\%$) of SDSS main sample galaxies(MGs) have not
been targeted with spectroscopy due to the the fiber collision effect. These
galaxies have been compiled into the input catalog of the LAMOST extra-galactic
survey and named as the complementary galaxy sample. In this paper, we
introduce the project and the status of the spectroscopies of the complementary
galaxies in the first two years of the LAMOST spectral survey(till Sep. of
2014). Moreover, we present a sample of 1,102 galaxy pairs identified from the
LAMOST complementary galaxies and SDSS MGs, which are defined as that the two
members have a projected distance smaller than $100\kpc$ and the recessional
velocity difference smaller than 500 $\kms$. Compared with the SDSS only
selected galaxy pairs, the LAMOST-SDSS pairs take the advantages of not being
biased toward large separations and therefor play as a useful supplement to the
statistical studies of galaxy interaction and galaxy merging.}


\authorrunning{Shiyin Shen et al. } 
\titlerunning{A sample of galaxy pairs } 
\maketitle

\section{Introduction} \label{sect:intro}

In the standard hierarchical structure formation model, galaxies are built up
through merging processes. Numerical simulations show that the galaxy mergers 
can trigger the star burst, feed the central super-massive black hole and
transform the galaxy morphology\citep{Springel05}. These different physical
processes take places at different stages of galaxy merging. At early stage, as
two galaxies approaching, they start to have interactions on each other out to
a distance of about 100 kpc. After the first passage, the galaxies start to
show strong tidal tails and undergo star bursts. After few times of passages,
the galaxies quickly evolve into final coalescence. The whole time scale of the
merging progress takes bout $1-2$ Gyr\citep{Torrey12}.

In observation, the processes of galaxy merging have been probed by statistical
studies of galaxy pairs as function of their separation, stellar mass,
morphology, mass ratio and many other
parameters\citep[e.g.][]{Nikolic04,Ellison08,Ellison13}. In such studies, a
large and unbiased sample of galaxy pairs is crucial. By far, the largest low
redshift galaxy
pair sample is identified from the main galaxy(MG) sample of the Sloan Digital
Sky Survey\citep[SDSS,][]{York00}, which is a spectroscopic survey of a
magnitude-limited sample down to $r<17.77$\citep{EDR}. The spectroscopic
completeness of the MG sample in SDSS is quite high \citep[$\sim
90\%$,][]{Hogg04}.
Based on the spectroscopic MGs in the final data release of the SDSS legacy 
survey\citep[Data Release Seven, DR7][]{DR7} , the number of galaxy pairs
 is over 10,000\citep[e.g.][]{Ellison11}.  Despite of this large number, the
galaxy pair sample identified from the spectroscopic MGs is far from complete.
The incompleteness of the galaxy pairs is mainly caused by the fiber collision
effect in SDSS, which is a minimum separation of 55 arcsec between any two
fibers for any given spectroscopic plate. As a result, the completeness of the
galaxy pairs identified from the spectroscopic MGs is estimated to be only about 
35 percent\citep{Patton08}. In other words, the SDSS missed galaxies caused by
 the fiber collision have a very high probability being in galaxy pairs.
Therefore, spectroscopic targeting of the SDSS missed galaxies is an efficient
way to identify new galaxy pairs. Only if all these SDSS missed galaxies
could be targeted by a new spectroscopic survey,  a complete
and unbiased galaxy pair sample could be finally made. More importantly, such a
sample would be a benchmark on the studies of  small-scale environmental effect
 of low redshift galaxies.

In this manuscript, we introduce the project of observing the SDSS missed MGs
with the Guo Shou Jing Telescope (also named as the Large Sky Area
Multi-Object Fiber Spectroscopic Telescope - LAMOST)\footnote{http://www.lamost.org}
and present its early result: a new
sample of galaxy pairs. This paper is organized as follows. In section 2, we
introduce the project of the spectroscopic observation of the SDSS missed MGs
with LAMOST. In section 3, we present a new galaxy pair sample using  new
redshifts from LAMOST survey. Finally, we make short discussions and give
summary in Section 4.

\section{LAMOST survey: complementary galaxy sample}
\label{sect:Obs}

LAMOST is a special quasi-meridian reflecting Schmidt
telescope located at Xinglong Station of National Astronomical Observatory of
China. The design of LAMOST provides  an effective aperture about 4 meters, a
diameter of $\sim 5^\circ$ field of view  and a
spectroscopy system with $\sim 4000$ fibers\citep{Wang96,Su04,Cui12}. After
about one year of pilot survey \citep{pilot}, the LAMOST regular spectral
survey has started from  September of 2012 and would last for 5 five years. 
An overview of the LAMOST  spectral survey can be seen in \citet{Zhao12}. 
The LAMOST regular survey mainly focuses on Galactic
stars, but also includes a significant fraction of extragalactic
objects\citep[e.g.][]{Huo13,Shi14}. One of the sample of the extragalactic
sources is the SDSS missed MGs , and is named as the
complementary galaxy sample in the LAMOST survey\citep{LDR1}. 

\subsection{the complementary galaxy sample}

\begin{figure}[htb]
\centering
\includegraphics[width=120mm]{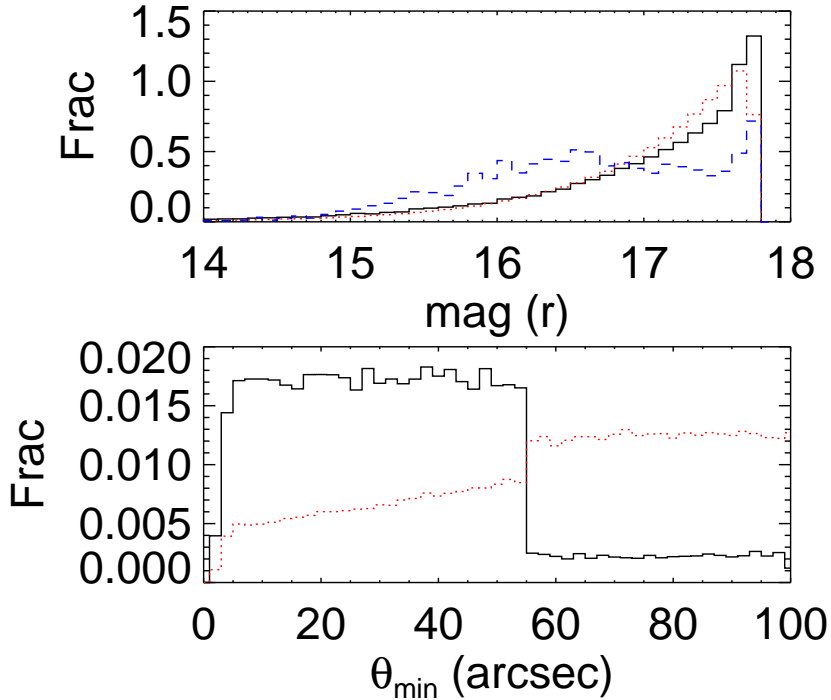}
\caption{The  SDSS spectroscopic MGs and complementary galaxy sample.
Top:  the $r$ band Petrosian magnitude distribution. The solid and
dotted histograms show the distributions of the complementary galaxies and SDSS spectroscopic
MGs respectively, while the dashed histogram shows the 3,456
complementary galaxies with spectra in LAMOST DR2(Section 2.2). Bottom: the angular separation to the
nearest SDSS MG(complementary galaxy: solid, SDSS spectroscopic MG: dotted). 
All the histograms have been normalized to  unit area. }
\label{Comp}
\end{figure}

The complementary galaxy sample is constructed from  the catalog archive server 
of the SDSS legacy survey, where all the galaxies with $r$ band Petrosian magnitude (Galactic reddening corrected)
brighter than $r=17.77$ and not yet with spectroscopic redshifts are selected. The footprint
of the complementary galaxies in the LAMOST survey is restricted in the north Galactic cap region
($ -10 < \delta < 60$ deg and $b> 0$ deg). \footnote{For the south Galactic
region, since the SDSS MGs are only located in three stripes, the LAMOST survey
includes another independent galaxy spectroscopic survey project, which aims to get the redshifts of of a
magnitude-limited sample down to $r<18$.} After that, we  remove a small fraction of galaxy targets
that might be contaminated by nearby bright stars using the spherical polygon
masks in the NYU value added galaxy catalog \citep{VAGC}. The final number of the
complementary galaxy sample in the input catalog of the LAMOST spectral survey is
66,263.

In SDSS DR7, the number of the MGs that have been targeted with spectroscopy in
north Galactic cap is 639,428. That is to say, the fraction of SDSS missed MGs(66,263 of 705,691) is about 10
percent. In Fig. \ref{Comp}, we show the $r$ band Petrosian magnitude distributions of the
complementary galaxy sample and the SDSS spectroscopic MGs respectively. As can be seen, the
complementary galaxies have similar magnitude distribution as the SDSS
spectroscopic MGs, but are slightly biased toward faint galaxies. Most of the
SDSS missed MGs are due to the fiber collision effect, therefor they are  expected to be biased
toward the high density region. To show this effect quantitatively, we match each complementary galaxy to the
global MG sample and obtain the projected distance $\theta_{\rm min}$(in unit of
arcsec) to its nearest neighbor. For comparison, we also calculate $\theta_{\rm
min}$ for each SDSS spectroscopic MG. The two distributions of $\theta_{\rm min}$ are plotted  in the bottom
panel of Fig. \ref{Comp}. Strong biases of two $\theta_{\rm min}$
distributions at 55 arcsec are clearly seen. This result not only shows that the
complementary galaxies are biased to the high density environment, but also
implies that the galaxy pair sample identified from the SDSS spectroscopic
sample alone is far from complete\citep{Patton08}.

\begin{figure}
\centering
\includegraphics[width=120mm]{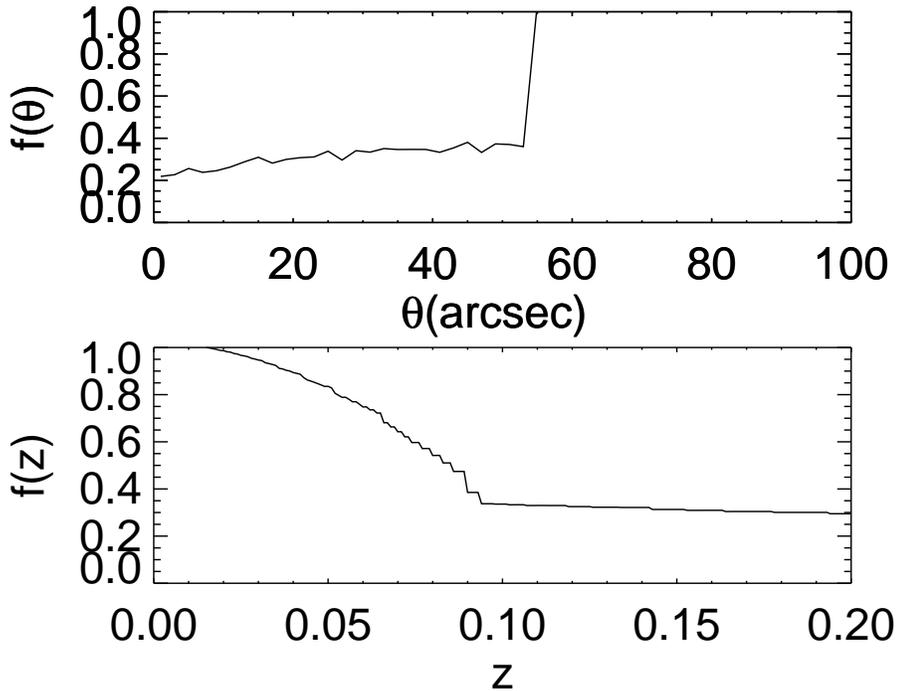}
\caption{The completeness of galaxy pairs in SDSS spectroscopic MGs. The top
panel shows the completeness as function of the angular separation of the pair
members. The bottom panel shows the completeness of the galaxy pairs (defined as
these with projected distance$\rp < 100\kpc$) as function of redshift. }
\label{fcomp}
\end{figure}

To quantify the incompleteness of the galaxy pairs in SDSS, we select photometric 
galaxy pairs in the SDSS MG sample, which are defined as  the galaxies and their nearest 
neighbors  inside the radii of 100 arsec.  We match photometric pairs from  the SDSS spectroscopic MGs and from 
all the MGs respectively. The fraction of the galaxy pairs of these two samples 
gives us a estimation of the completeness of the galaxy pairs in SDSS spectroscopic MGs. We show the resulted
completeness as function of the pair angular separation in the top panel of Fig.
\ref{fcomp}. Again, we see that a strong incompleteness jumps at the angular
separation $\theta < 55$ arcsec, where the completeness is only about 30 percent
and even decreases with the decreasing of $\theta$. In the bottom panel, we show
the completeness of the galaxy pairs as function of the redshift. Here, we have
defined the galaxy pairs as these with  projected separation  $\rp < 100
\kpc$(see Section 3) and assumed that the separation follows a random
distribution. In this case, the galaxy pairs in SDSS spectroscopic MGs decreases
with increasing of redshift and reaches a plateau $\sim 30$ percent at $z>0.09$
where the 55 arcsec limit corresponds to a projected distance 100$\kpc$.
Since the peak redshift of the SDSS MGs is at $z\sim0.1$,
the global completeness of the galaxy pairs in SDSS spectroscopic MG sample is
less than 40 percent.

\subsection{LAMOST observation}

The  complementary galaxies are  mixed together with other LAMOST
targets(most of them are Galactic stars) and then compiled into the LAMOST
survey plates. In each plate, the number density of the complementary galaxies is very low,
which are therefore assigned fibers with higher priority than stars. In the
LAMOST survey, the input sources are tiled into three different types of plates, 
the bright(B), medium(M) and faint(F) plates, which are designed to reach the
average signal-to-noise ratio ($S/N\sim10$) for the objects down to the
magnitude limits  $r<16.5,r<17.8$ and $r<18.5$ respectively. Most of the
complementary galaxies have their magnitudes in the range $16.5<r<17.8$ 
(Fig. \ref{Comp}) and therefor are mainly tiled into the M plates. On the other hand, due to the
limited number of the dark nights, the observing time allocated for the M
plates is quite few. Because of that, only a small fraction of the complementary
galaxy sample have been targeted yet.

The spectroscopic data used in this study is from the LAMOST Data Release 2
(DR2) , i.e the data till June 2014 . In LAMOST DR2, there are 3,456 complementary
galaxies that have been targeted with spectroscopy and with spectra released.
The magnitude distribution of these 3,456 LAMOST targeted galaxies is plotted as
the dashed histogram in the top panel of Fig. \ref{Comp}. Compared with the
input  of the 66,263 complementary galaxies, the LAMOST targets are
evidently biased toward  bright galaxies. There are two reasons for this
bias. First, some of the bright complementary galaxies($r<16.5$) are 
compiled into the target list of LAMOST B plates, which far outnumbers the M plates.
The other reason for the bias against the faint galaxies is the
failure of the spectroscopies because of their low $S/N$(see next section).

\subsection{redshift measurements}

\begin{figure}
\centering
\includegraphics[width=120mm]{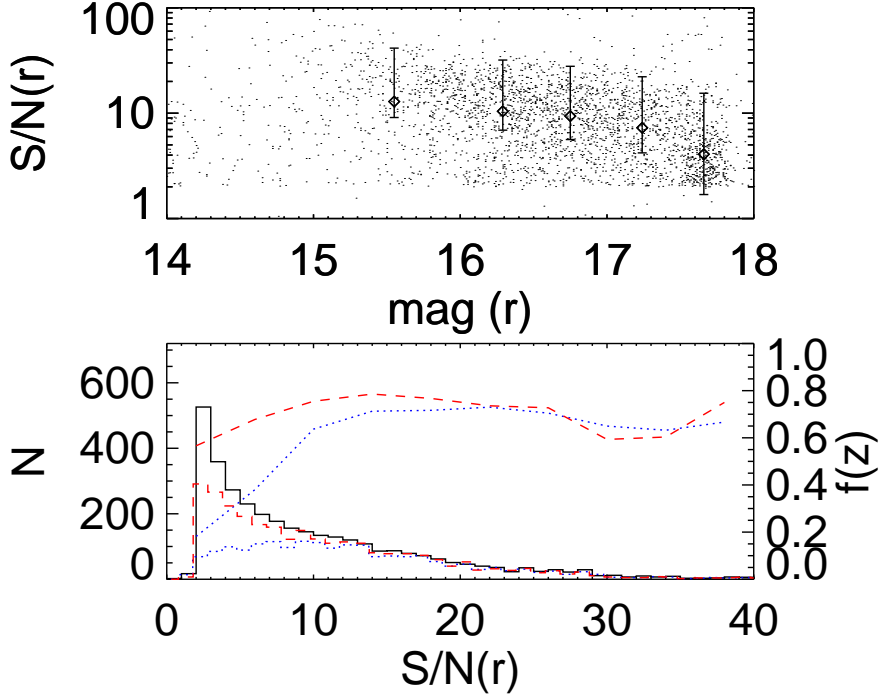}
\caption{The signal-to-noise ratio($S/N$)  of the LAMOST spectra of the complementary
galaxies. The top panel shows the magnitude as function  of $S/N$, while  the bottom panel shows the 
histograms of $S/N$.  Top: the small dots represent all 3,456 complementary galaxies. The squares 
with error bars show the median and 16/84 percentiles of the $S/N$ distribution in magnitude bins. 
Bottom: the solid histogram shows all the 3,456 complementary galaxies, while the
dashed and dotted histograms show the sub-samples of galaxies with redshifts
measured from the PCA algorithm and form the LAMOST 1-D pipeline respectively.
The fractions of galaxies with redshift measured(right $y$-axis) for the spectra
at different $S/N$ are shown by the dotted line(PCA algorithm) and dashed
line(1-D pipeline).}
\label{SNR}
\end{figure}

In the catalog of LAMOST DR2, only 1,951 of the 3456 complementary galaxies have redshifts
measured from the LAMOST 1-D pipeline and published in the LAMOST catalog.
The failure of the redshift measurements is mainly due to the low $S/N$ of the LAMOST
spectra of the faint galaxies($r>16.5$). We show the mean $S/N$ in the $r$-band wavelength range
against the $r$-band magnitude of the 3,456 LAMOST targeted galaxies in the top panel of Fig. \ref{SNR}.
The distribution of the $S/N$ is shown in the bottom panel.
As can be seen, when $r>16.5$, the median $S/N$ of the LAMOST spectra becomes lower than 10,
 which makes the number of the spectra with LAMOST catalog $z$ decreases significantly
 (see dotted histogram in the bottom panel of Fig. \ref{SNR}).
 
To further improve the successful rate of the redshift measurement, we developed
an independent PCA redshift measurement algorithm following the pipeline used
for the `CMASS' galaxies in the Baryon Oscillation Spectroscopic Survey of the
SDSS III\citep{Bolton12}. The detail of the pipeline will be presented in an
upcoming paper. Here, we outline the basic routines in Appendix A. With this
algorithm, we obtained 2,796 redshift measurements.  In Fig. \ref{SNR},  
the  dashed line shows the fraction of the spectra with PCA
redshifts measured. It is clearly that our new PCA algorithm makes a significant
improvement on the redshift measurement of the LAMOST spectra, especially at the
low $S/N$ end.

\begin{figure}
\centering \includegraphics[width=120mm]{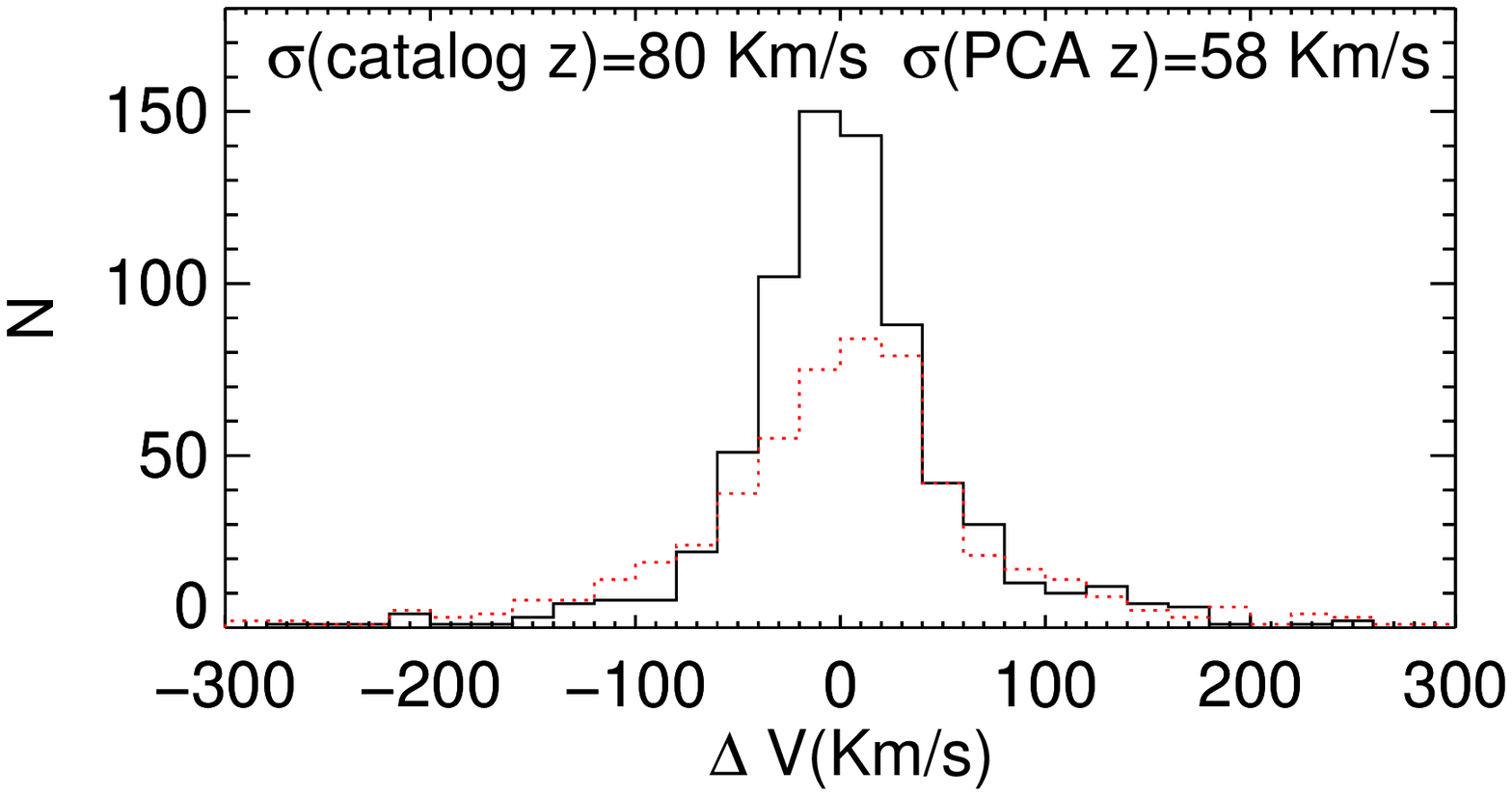}
\caption{Number histogram of the redshift differences (in terms of recessional
velocity) between the LAMOST and SDSS measurements. The solid histogram shows the differences
between the SDSS redshifts and our PCA redshift measurements, while the dotted histogram represents the 
differences between the SDSS redshifts and and LAMOST 1-D redshifts. }
 \label{LSz}
\end{figure}

To  quantify the redshift measurements from the LAMOST 1-D pipeline and
our PCA algorithm, we compare their redshifts with other
independent measurements. After the SDSS legacy survey, some of the SDSS missed
MGs had been targeted by the BOSS spectroscopy in SDSS III\citep{BOSS}
\footnote{These galaxies typically have 
photo flag BRIGHT$\_$GAL in SDSS III.}. We matched the 3,456 complementary
galaxies with SDSS DR12 spectroscopic catalog and obtained 1,056 matches.   
For these 1,056 galaxies, the LAMOST catalog lists 604 redshifts 
while our PCA algorithm provides 923 redshift measurements. In Fig. \ref{LSz}, we show the
histograms of the differences of the redshifts (in
terms of recessional velocity difference $\Delta V$) of these galaxies with both
LAMOST and SDSS redshifts. The LAOMST redshifts from 1-D pipeline and PCA algorithm both show
good consistences with SDSS values.
For the LAMOST catalog $z$, the standard deviation of $\Delta V$ is about 80$\kms$.
For the PCA $z$, the scatter of $\Delta V$ is even smaller, $\sim 58\kms$. Given the
better consistence with the SDSS redshifts of the  PCA redshift measurements, we
use the PCA redshifts for these galaxies with both PCA redshifts and
catalog redshifs. For the 1,056 galaxies with SDSS redshifts, we take their
redshifts from SDSS catalog. As we will show in the next section, the criterion
of the velocity difference we adopted to identify galaxy pairs is $\dv < 500
\kms$. Therefore, the scatter between the SDSS and LAMOST redshift measurements
 have few impacts on the pair identification.

\section{the  galaxy pair sample}

\begin{figure}
\centering \includegraphics[width=120mm]{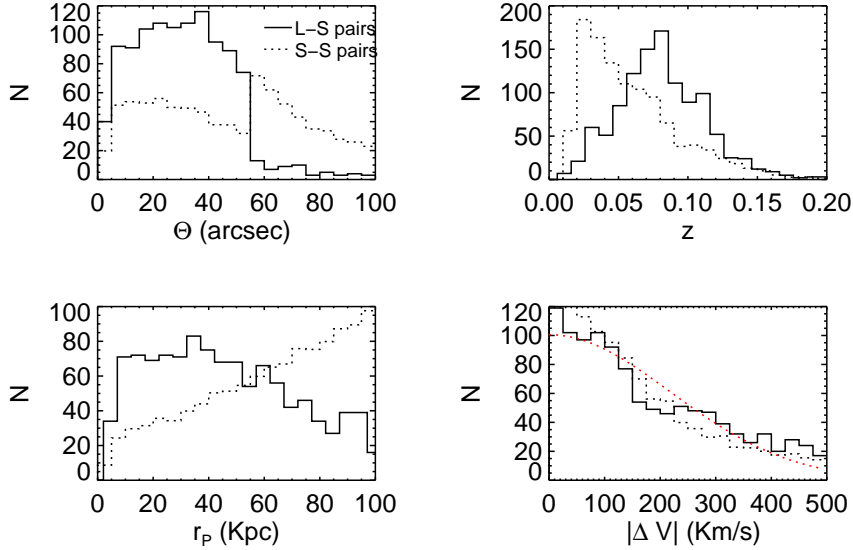}
\caption{Basic statistical properties of the 1,102 galaxy pairs identified
from the LAMOST complementary galaxies and SDSS MGs(solid histograms). The
statistical properties of the SDSS selected galaxy pairs are plotted as the dotted histograms
in each panel for comparison (normalized to 1,102). The top left, top right, bottom left and
bottom right panels
show the histograms of the angular separation, redshift, projected distance and
recessional velocity differences respectively. In the bottom right panel, the
dotted curve shows a Gaussian
distribution function with a standard deviation $218 \kms$.}
\label{hBasic}
\end{figure}

In this section, we combine the  redshifts of the complementary galaxies
with the SDSS spectroscopic MGs to identify  new galaxy pairs. For the 3,456 complementary
galaxies, we have
obtained 3,137 redshifts. Among them, 1,056 redshifts come from SDSS DR12, 1906
from PCA algorithm, 175 from
LAMOST 1-D pipeline.

In observation, a galaxy pair is typically defined from the projected
distance $\rp$ and recessional velocity difference $\dv$ of two neighboring
galaxies.
However, there is no consensus on  the critical values of  $\rp$ and $\dv$.
For example, both based on SDSS DR7, \citet{Liu11} defined an AGN pair sample
with
$\dv < 600 \kms$ and  $\rp < 100\kpc$, whereas \citet{Patton11} searched
galaxy pairs using $\dv < 1,000\kms$ and $\rp < 80\kpc$\citep[see
also][]{Scudder12,Maria15}.
For the projected distance $\rp$,
there are evidences that galaxies show interactions on their neighbors at $\rp >
80\kpc$ \citep{Scudder12}.
In this study, we set a critical value $\rp < 100 \kpc$. For $\Delta V$, a large critical
value (for example, $\dv<1,000 \kms$) might introduce a significant
fraction of contamination from the high density environment(e.g. galaxy groups
and clusters). In this study, we select galaxy pairs using
$\dv < 500 \kms$ (see Appendix B for more discussions).

We match the redshifts of 3,137 complementary galaxies with the
spectroscopic MGs in SDSS DR7 using the criteria
$\rp< 100 \kpc$ and $\dv< 500 \kms$ and obtain 1,141 galaxy pair candidates. In
a few cases, a complementary galaxy may have more than one SDSS spectroscopic MGs matched. In this
case, we choose the galaxy with the smallest $\rp$ as the matched pair member
and mark this pair  being in a multiple system. We will come to the multiple
systems in Section 3.1. Moreover, to have a better quality control to the pair 
sample, we make visual inspections on the SDSS images for all the pair
candidates. In a few cases, the imperfect SDSS pipeline de-blends big galaxies
into several small children.
This effect results in 39 fake pairs. Thus, our final sample include 1,102 galaxy
pairs.

We show the histograms of the $\rp$ and $\dv$ of the final 1,102 galaxy pairs in
the bottom two panels of Fig. \ref{hBasic}, where the distributions of
the angular separation $\Theta$ (in acrsec) between the pair members and their
average redshifts are shown in the top two panels. To have a better understanding 
of the statistical properties of the new pair sample, we show the distributions of the SDSS only pairs as the  dotted 
histograms in each panel of Fig. \ref{hBasic} for comparison. The SDSS only
pairs are selected from the SDSS DR7 group catalog of \citet{Yang08} using the same criteria 
above. The number of the SDSS pairs is 16,973. Because of its large number,
we have not made visual inspections on this sample.  However,
according to the fact that there are only 39 of 1,141 LAMOST-SDSS pairs are fake, 
we expect the impact of the fake pairs to its statistical properties should be
quite small. For convenience, we abbreviate the LAMOST-SDSS pairs as the LS pairs
and the SDSS only  pairs  as SS pairs below.

As can be seen from Fig. \ref{hBasic}, except the distribution of $\dv$, the LS
pairs show significant difference from the SS pairs. The
angular separation of the SS pairs is clearly biased toward large values($\Theta > 55$
arcsec, top left panel). This, as we mentioned, is because of the 55 arcsec
fiber collision effect in SDSS. The SS pairs are also biased toward lower
redshifts(top right panel), which again is because of the fiber collision
effect. Although the low $z$ selection effect compensates the high
$\Theta$ bias, as a combination, the SS pairs are still biased toward large
separation ($\rp$) ones (bottom left panel).  Compared with the SS pair
sample, our new LS pairs only increase the number of pairs by a few percent ($\sim 7\%$).
However,  for the close pairs with more
significant interaction$(\rp < 30\kpc)$,  the number of the
LS pairs enlarges the SS pairs by more than 14 percent(358 versus 2527). Therefor, the LS pair sample
 could be a useful supplement to the current SS pair sample, especially  for the close pairs, 
 which are valuable in the statistical studies of galaxy interaction and merging . 

For the velocity difference (bottom right panel), the standard deviation of the LS and SS pairs is 218 and 198
$\kms$ respectively.  Considering
the fact that the dispersion of the redshift differences between the LAMOST and
SDSS measurements could be as large as $\sim 80\kms$(Fig. \ref{LSz}), these two results are
consistent with each other. In this panel, we also plot a Gaussian
distribution with standard deviation $\sigma=218\kms$ for comparison. As we can
see, the distribution of $\dv$ of LS pairs deviates from the Gaussian
function, especially at the  tails (see Appendix B for a more detailed
discussion).

The catalog of the LS pairs are listed in Table 1. Besides  basic
parameters(e.g. Ra, Dec, redshift) for each pair member, we
also list two extra flags($\Mflag$ and $\Oflag$) for each pair, which characterizes the
multiplicity and the image overlapping of the pair members respectively. 
 In the following two sub-sections, we make brief
descriptions on these two flags.

\begin{table}
\begin{center}
\caption{The catalog of galaxy pairs identified from the LAMOST complementary
galaxies and SDSS MGs. For each pair, $\alpha_1,\delta_1,z_1$ are the Right Accession,
Declination,  redshift of the LAMOST complementary galaxy, while
$\alpha_2,\delta_2,z_2$ are those of the SDSS galaxy.
$\Mflag$ and $\Oflag$ are the multiplicity and overlapping flags (see Section
3.1 and 3.2 for detail). The table is sorted in ascending order of $\alpha_1$. The
complete table is available on-line.}

\label{Tab_pair}
\begin{tabular}{lllllllll}
\hline\noalign{\smallskip}
ID & $\alpha_1$[deg] & $\delta_1$[deg] & $z_1$ & $\alpha_2$[deg] &
$\delta_2$[deg] & $z_2$ & $\Mflag$ & $\Oflag$ \\
1 & 112.23653 & 36.91987 & 0.06006 & 112.24269 & 36.91610 & 0.05988 & 0 & 0 \\
2 & 114.54001 & 28.13682 & 0.07852 & 114.54659 & 28.12799 & 0.07942 & 0 & 0\\
3 & 116.04450 & 23.99016 & 0.07545 & 116.05284 & 23.99939 & 0.07533 & 0 & 0\\
4 & 116.46624 & 26.47179 & 0.12311 & 116.46654 & 26.47719 & 0.12373 & 0 & 0\\
5 & 119.86285 & 23.97182 & 0.09218 & 119.85204 & 23.98273 & 0.09318 & 0 & 0\\
6 & 120.09847 & 39.83033 & 0.01320 & 120.17009 & 39.87050 & 0.01326 & 1 & 0\\
7 & 120.40500 & 15.70968 & 0.01545 & 120.36157 & 15.74989 & 0.01637 & 1 & 0\\
8 & 120.83544 & 23.96435 & 0.05784 & 120.84409 & 23.96933 & 0.05725 & 0 & 0\\
9 & 120.89831 & 28.54465 & 0.14178 & 120.89262 & 28.55000 & 0.14096 & 0 & 0\\
10 & 121.02204 & 31.44029 & 0.07302 & 121.03594 & 31.43697 & 0.07303 & 0 & 0\\
11 & 121.37741 & 22.13299 & 0.13852 & 121.38057 & 22.12452 & 0.14003 & 0 & 0\\
12 & 123.44543 & 8.38181 & 0.11409 & 123.44723 & 8.38925 & 0.11324 & 0 & 0\\
13 & 123.98402 & 8.28538 & 0.14295 & 123.97318 & 8.28571 & 0.14438 & 0 & 0\\
14 & 124.05299 & 3.85966 & 0.08668 & 124.05083 & 3.85807 & 0.08777 & 0 & 1\\
15 & 124.46973 & 7.57755 & 0.12465 & 124.46754 & 7.57477 & 0.12484 & 0 & 0\\

\hline
\end{tabular}
\end{center}
\end{table}

\subsection{multiple system}
\begin{figure}
\centering \includegraphics[width=120mm]{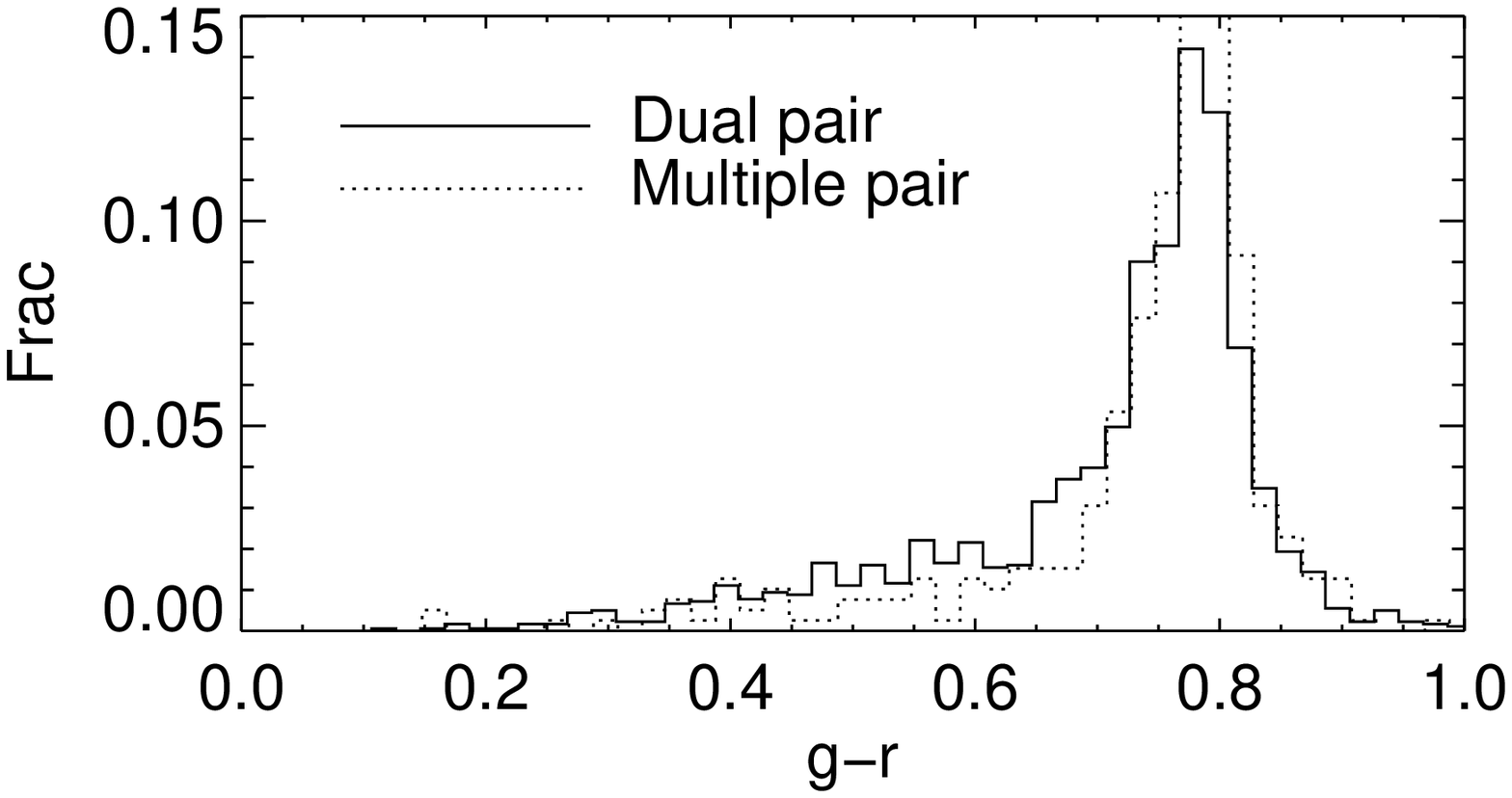}
\caption{Distribution of the the $g-r$ color of the pair members. The solid and
dotted histograms (normalized to unit area) show the galaxies in the pairs with
$\Mflag=0$ (dual) and $\Mflag=1$
(multiple) respectively.}
\label{Mflag}
\end{figure}

Our galaxy pair sample is defined from two simple observational criteria
 ($\rp <100 \kpc$ and $\dv < 500 \kms$) .  Besides the pair members, the other
neighboring galaxies have not been taken into consideration. However, to study the galaxy
interaction and galaxy merging using galaxy pairs, it is better to refine a
sample of $physical$ pairs, where the interaction between two pair members
overtakes the effects from  other neighbors\citep[e.g.][]{Maria15}. For this
purpose, we mark the galaxy pairs in multiple system with a multiple flag `$\Mflag$'. 
Our aim is to remind that these pairs may not be suitable for studying galaxy interactions by
only considering their members.

We define a galaxy pair in multiple system when either of its member has another
main galaxy neighbor($r<17.77$) that also satisfies the pair definition($\rp < 100 \kpc$ and 
$\dv < 500\kms$) . Within current data, 197 of the 1,102 LS pairs are found in
multiple systems. These pairs are marked with flag $\Mflag = 1$ in Table 1 and
noted as `multiple' pairs below. For comparison, the pairs with $\Mflag = 0$ are
noted as `dual'  pairs.

To check the possible distinction between the multiple and dual
pairs, we compare the $g-r$ color(taking from the SDSS model magnitudes and
with $K$-correction applied) distributions of their  members in Fig.
\ref{Mflag}. The galaxies in multiple pairs are averagely redder than in dual ones. 
In specific, the red galaxy fraction($g-r>0.7$) in the dual
and multiple pairs are 0.52 and 0.66 respectively. This color bias is caused by
the fact that the multiple pairs are biased toward high density environment (i.e. galaxy groups
and clusters).

We remind that the $\Mflag$ set for the multiple systems is quite a preliminary parameter.
 Since the pairs are selected from
a magnitude-limited sample, both of the pair sample
and the multiplicity flag have  strong redshift dependence. For example, a
triplet system with $M_{{\rm r},i}=-20,-21,-22$ mag would be identified as a
triplet, a pair and a single galaxy in the SDSS MG sample ($r<17.77$) at
redshift $z=0.05,0.1,0.15$ respectively.
On the other hand, the multiplicity flag is also not complete, which is because
of the high incompleteness of the SDSS spectroscopic MGs at small
scales and also because of the small fraction of  the complementary galaxies been
 targeted by LAMOST yet. For the dual pairs with $\Mflag=0$, they are still
possibly in multiple systems, i.e. having another companion galaxy but without
redshift measured yet.\footnote{The pair with ID=52 in Table 1 is actually an
isolated and compact galaxy triplet, which is found to be a
triplet candidate during our visual inspection and then spectroscopically
confirmed by a follow-up observation with the 2.16m telescope
in the Xinglong Station(Feng et al. , 2015) .} That is to say, the fraction of
the pairs in multiple
system (197 of 1102) is actually a lower limit. Given the significant fraction
of galaxy pairs in multiple system, the studies of galaxy interactions with
galaxy pairs should take the multiplicity of the galaxy systems into
consideration(Shen et al. 2015, in preparation).

\subsection{overlapping pairs}
\begin{figure}
\centering \includegraphics[width=120mm]{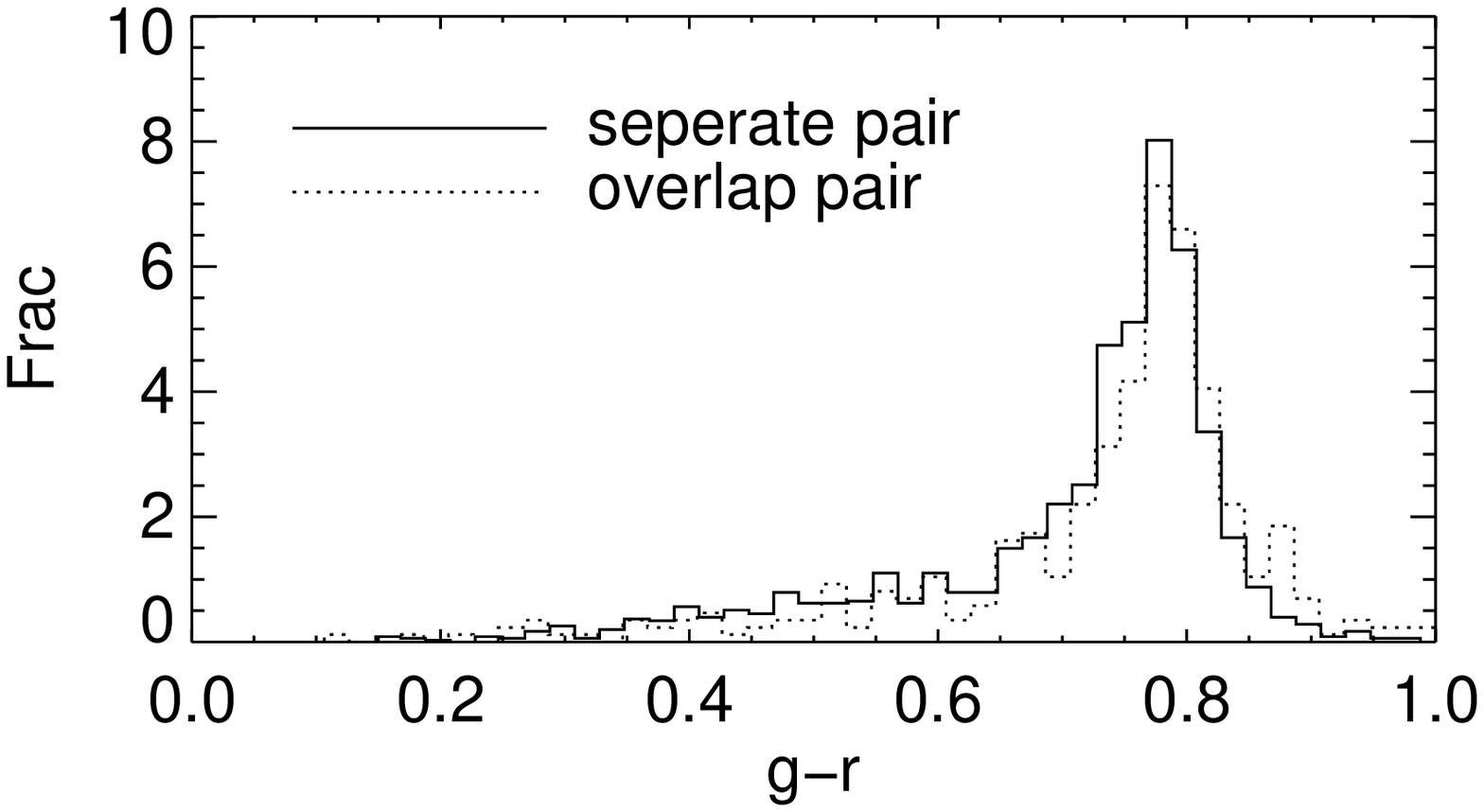}
\caption{Distribution of the $g-r$ color of the pair members. The solid and
dotted
histograms(normalized to unit area) show the galaxies in the pairs with
$\Oflag=0$ and $\Oflag=1$
respectively.}
\label{Oflag}
\end{figure}

The luminosity/mass ratio of the pair members plays an important role in  galaxy
merging\citep{Jiang14}.  However,  photometry of a very close galaxy
 pair is a nontrivial task\citep{Simard11}. During the visual inspection
of the galaxy pairs, we noticed some of the galaxy pairs whose images of their
members overlap each other significantly. We mark such pairs($N=216$) with a
flag $\Oflag = 1$.  For these overlapping pairs, the uncertainties of their photometry might
introduce uncertainties and possible biases in the estimation of the stellar
mass and mass ratio.

We show the $g-r$ color distribution of the
overlapping pairs as the dotted histogram in Fig. \ref{Oflag}, where the
distribution of the other pairs is shown as the solid histogram for comparison.
As can be seen, although the overlapping pairs have much closer projected
distance, their colors are biased toward redder values. The fraction of red
galaxies($g-r>0.70$) in
the overlapping pairs is 62 percent, which is significantly higher than the
fraction 0.52 for the non-overlapping pairs.
Unless there are big uncertainties in the photometry of these overlapping pairs, it is hard to
explain the color bias we see in Fig. \ref{Oflag}. Actually, \citet{Patton11}
have already shown that the poor photometry of SDSS official pipeline is largely responsible
for the suspicious and large fraction of
of extreme red galaxies(e.g. $g-r>0.9$) in the very close galaxy pairs. Therefore, to
get a better estimation of the mass ratio of these overlapping pairs, more detailed
photometry is required\citep[e.g.][]{Simard11}. Here, we set this flag for a
caution and leave the detailed photometry for a future work.

\section{Conclusion}

In this paper, we presented the project of the spectroscopic survey of the SDSS
missed MGs($r<17.77$) using LAMOST. The SDSS missed MGs are named as the
complementary galaxy sample in the LAMOST survey. In the first two years of the
LAMOST survey, due to the limited survey time of the medium(M) plates, only a
small fraction (3,456 of 66,263) of the complementary galaxies had obtained
LAMOST spectra.  The majority of these spectra have quite low $S/N$, 
which are mostly due to the poor seeing condition.
We developed a PCA algorithm to improve the redshift measurements of these low
$S/N$ spectra. Together with the SDSS DR12 match and LAMOST 1-D pipeline results, we finally
obtained 3,137 redshifts of the 3,456 complementary galaxies.

Considering the fact that the SDSS missed galaxies are mainly caused by the
fiber collision effect, the spectroscopy of the complementary galaxies has a
great potential in identifying new galaxy pairs. We present such a catalog of
galaxy pairs identified from the first two years data of the LAMOST survey. From
the redshifts of 3,137 complementary galaxies, we obtained a sample of 1,102 galaxy
pairs after a careful visual inspection.  Compared with the galaxy pairs selected from SDSS
data only, our pair sample includes a  larger fraction of close pairs($\rp < 30\kpc$). Because of such advantages, 
our sample increased the current SDSS close pairs  sample by  about an 
amount of $\sim15$ percent.  In common with other studies, our pairs are selected
using two simple observational criteria($\rp$ and $\dv$) . Whether they are physical bounding systems have
not been taken into consideration. We find that at least  $\sim20$ percent of the pairs are actually located in multiple
systems(($\Mflag =1$) ). Moreover, during the visual inspection, we find, for about 20 percent the pairs, the images of
 their members overlap each other. Therefor, the photometry and stellar mass estimation of the galaxies in these overlapping
pairs($\Oflag =1$) should be in caution. 

With the acceleration of the LAMOST survey on the complementary galaxies in the
current and future season(from Sep. of 2014), we expect that the completeness of 
the galaxy pairs will be further improved. Once the LAMOST survey had finished
the spectroscopies of the complementary galaxy sample, the great progresses on
the studies of galaxy pair, galaxy interactions, galaxy merging and the small-scale
environmental effects of galaxies would be expected. 

\begin{acknowledgements}

Guoshoujing Telescope (the Large Sky Area Multi-Object Fiber Spectroscopic
Telescope, LAMOST) is a National Major Scientific Project built by the Chinese Academy of
Sciences. Funding for the project has been provided by the National Development and Reform
Commission. LAMOST is operated and managed by the National Astronomical Observatories,
Chinese Academy of Sciences.

This work is supported by the ``973 Program'' 2014 CB845705, Strategic Priority
Research Program ``The Emergence of Cosmological Structures'' of the Chinese Academy of Sciences
(CAS; grant XDB09030200) and the National Natural Science Foundation of China (NSFC) with the Project Number of 11573050 and 11433003.
\end{acknowledgements}

\label{lastpage}
\appendix

\section{redshift measurements with PCA eigen-templates}

We developed an independent algorithm to measure the redshifts of the LAMOST
complementary galaxies using PCA eigen-templates and chi-squared fitting.

The eigen-templates are derived from the 1,056 complementary galaxies with DR12
redshifts. In specific, we first divide 1,056 galaxies into 20 $g-r$ color bins with similar numbers.
Then, using the SDSS DR12 redshifts, we stack their
LAMOST spectra and build the LAMOST composite spectra for galaxies in each $g-r$ bin.
We make a PCA decomposition for these 20 composite spectra and  find that the
first 4 eigenvectors can recover most of the spectroscopic features.

We explore the redshift of each galaxy in the range $0.005<z<0.5$ by taking
trial values that are moved with steps of each spectrum pixels.
For each trial redshift, we fit the observed spectrum with the error-weighted
least-square linear combination of the four `eigen-spectra` and
a four-order poly-nominal. The poly-nominal is introduced to compensate the 
calibration of the LAMOST spectrum.  The reduced $\chi^2$ value(the resulted $\chi^2$
divided by the number of fitting pixels) for each trial redshift define a
$\chi^2(z)$ curve in the probed redshift range. The best redshift estimation is
then defined by the minimal of the $\chi^2(z)$ curve. The error is evaluated at
the location where the $\chi^2$ is increased by one at each side of the minimum
values. During the fitting, besides the pixels with ANDMASK set, we also have masked
the pixels at the wavelengths where the sky-subtraction residuals are 3 times
more than the average noise.

\section{The pair-wise peculiar velocity distribution of galaxy pairs}

The distribution of the recessional velocity difference between the galaxy pair
members, also known as the
pair-wise peculiar velocity distribution function(PVDF), plays a very important
role in the large scale structure studies.
On the non-linear scales($r<1$ Mpc), both observations and theoretical models
suggest an exponential
PVDF, which can be explained and approximated by a weighted integral of Gaussian
distributions
 of subunits\citep[e.g.][]{Diaferio96,Sheth96}.
However, the PVDF of the pairs in our analysis is on very small scale, even
smaller than the
viral radius of galaxies ($r<100$ kpc), which has not drawn much attention in
related studies.

\begin{figure}
\centering \includegraphics[width=120mm]{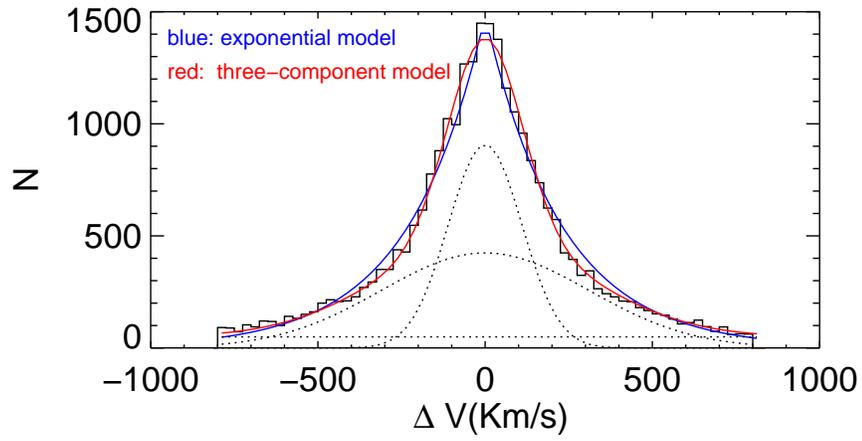}
\caption{Recessional velocity difference $\Delta V$ distribution of the galaxy
pairs in SDSS. The best exponential model fitting is shown as the blue line. The best
three-component model fitting is shown as the red line, where the contributions of each component are shown as
the dotted lines.}
\label{dV}
\end{figure}

We show the pair-wise peculiar velocity($\Delta V$) distribution of the SDSS
selected pairs in Fig.
\ref{dV}. The SDSS pairs are selected with criteria $\rp<100$ kpc and $\dv <
800\kms$. As we
can see, the $\Delta V$ distribution shows long tails out to $\pm 800 \kms$ and
can not be fitted well
by a Gaussian function. We first fit the observed $\Delta V$ distribution with
the usually adopted exponential profile,

\begin{equation}
f(\Delta V)=\frac{1}{\sqrt{2}\sigma} {\rm exp} \left(-\frac{\sqrt{2}\Delta
V}{\sigma} \right)\,,
\end{equation}

where $\sigma$ characterizes the pair-wise peculiar velocity dispersion.
The best fitting of the exponential profile has $\sigma=324\kms$ and is shown as
the blue line in Fig. \ref{dV}. Although better than the Gaussian profile fitting, the exponential profile still
can not fit well the fine structure of the observed $\Delta V$ distribution, especially at $\dv > 200\kms$ ranges.

Alternatively, as a preliminary test, we fit the observed $\Delta V$
distribution with a multi-component model.
The observed $\Delta V$ distribution might be contributed by different
components and each component corresponds to different physical circumstances. The ideal
case of a galaxy pair is that the pair members form a gravitational bound system. In this
case, $\Delta V$ represents the orbital velocity of the pair members. A more
common case of the observed galaxy pair is that the galaxy pairs have other
close neighbors. That is to say, the pair members and their neighbors together
locate in a bigger galaxy system, e.g. galaxy groups and clusters. In
this case, the pair members orbit individually in the gravitational potential of 
the host halo and $\Delta V$ represents its  velocity dispersion. Finally,
$\Delta V$ may  be dominated by the difference of the Hubble flow. In this
case, the radial distance of the members ($\sim$ Mpc) is much larger than
the tangential distance ($<100$ kpc). That is to say, the `pair' members
actually have few physical interactions on each other and such pairs could be considered as
contaminations from projection. Motivated by above scenario, we fit the observed
$\Delta V$ distribution with three components, one narrow Gaussian profile(orbital velocities of  ideal
pairs), one broad Gaussian profile(velocity dispersion of host halos) and a flat 
component(projection contamination). The best fitting of this three-component
model is shown as the solid line in Fig. \ref{dV}. As we can see, this model makes a pretty
good fitting to the observed $\Delta V$ distribution. The $\Delta V$ profiles of
the three components are  shown as the dotted lines in Fig. \ref{dV}. The narrow Gaussian
component has a standard deviation $\sim 110\kms$ and contributes about 40
percent of the galaxy
pairs with $\dv<800\kms$. The broad component has a deviation $\sim 300\kms$ and 
makes a half contribution. The constant component contributes the last 10
percent.

In Section 3.2, we select the LAMOST-SDSS galaxy pairs with criteria
$\rp <100$ kpc and $\dv < 500\kms$. Our motivation of choosing a smaller
critical value of $\dv$ is to reduce the contribution from  galaxy
groups and clusters, i.e. the broad component in Fig. \ref{dV}, while include
all
possible circumstances of ideal pairs. A more detailed study on the $\Delta V$
distribution
of the galaxy pairs in different environments is in preparation.

\end{document}